%% file: ms.tex
\documentclass[10pt,journal,compsoc]{IEEEtran}

\ifCLASSOPTIONcompsoc
  \usepackage[nocompress]{cite}
\else
  \usepackage{cite}
\fi

\usepackage{amsmath}
\usepackage{hyperref}

\usepackage[nameinlink,noabbrev]{cleveref}

\usepackage{algorithm2e}
\usepackage{blindtext}
\usepackage{colortbl}
\usepackage{svg}
\usepackage{hhline}
\usepackage{tikz,pgfplots}
\usepackage{verbatim}
\usepackage{tabularx}
\usepackage[caption=false]{subfig}
\usepackage{graphicx}
\usepackage{enumitem}
\usepackage{csquotes}
\usepackage{algpseudocode}
\usepackage{diagbox}
\usepackage{makecell}
\usepackage{threeparttable}
\usepackage{multirow}
\usepackage{balance}
\pgfplotsset{compat=1.17} 

\begin{document}
\title{mPSAuth: Privacy-Preserving and Scalable Authentication for Mobile Web Applications}

\author{David~Monschein~and~Oliver~P.~Waldhorst,~\IEEEmembership{Member,~IEEE}% <-this % stops a space
\IEEEcompsocitemizethanks{\IEEEcompsocthanksitem D. Monschein and Oliver P. Waldhorst are with the Data-centric Software Systems (DSS) Research Group at the Institute of Applied Research (IAF), Karlsruhe University of Applied Sciences, Karlsruhe, 76133 Germany\protect\\
E-mail: \{david.monschein,oliver.waldhorst\}@h-ka.de
}
\thanks{This work has been submitted to the IEEE for possible publication. Copyright may be transferred without notice, after which this version may no longer be accessible.}}

%\markboth{Journal of \LaTeX\ Class Files,~Vol.~18, No.~9, September~2020}%
%{How to Use the IEEEtran \LaTeX \ Templates}

\IEEEtitleabstractindextext{%
\input{sections/abstract}

}

\maketitle

\IEEEdisplaynontitleabstractindextext

\IEEEpeerreviewmaketitle

% 1. introduction
\input{sections/introduction}

% 2. foundations -> background and related work
\input{sections/foundations}

% 3. threat model and assumptions
\input{sections/threatmodel}

% 4. approach overall
\input{sections/approach/approach}

\input{sections/approach/overview}
\input{sections/approach/monitoring}
\input{sections/approach/privacy}
\input{sections/approach/detectors}

% 5. evaluation setup, evaluation results
\input{figures/evaluation/values}
\input{sections/evaluation/setup}
\input{sections/evaluation/results}

% 6. related work
\input{sections/relatedwork}

% 7. conclusion
\input{sections/conclusion}

% 8. acknowledgements
\section*{Acknowledgments}
This work was funded by the German Federal Ministry of Education and Research (BMBF), RefNr. 16KIS1142K.

\bibliographystyle{IEEEtran}
\bibliography{ms}

\end{document}

%% file: sections/abstract.tex
\begin{abstract}
As nowadays most web application requests originate from mobile devices, authentication of mobile users is essential in terms of security considerations. To this end, recent approaches rely on machine learning techniques to analyze various aspects of user behavior as a basis for authentication decisions. These approaches face two challenges: first, examining behavioral data raises significant privacy concerns, and second, approaches must scale to support a large number of users. Existing approaches do not address these challenges sufficiently. We propose mPSAuth, an approach for continuously tracking various data sources reflecting user behavior (e.g., touchscreen interactions, sensor data) and estimating the likelihood of the current user being legitimate based on machine learning techniques. With mPSAuth, both the authentication protocol and the machine learning models operate on homomorphically encrypted data to ensure the users' privacy. Furthermore, the number of machine learning models used by mPSAuth is independent of the number of users, thus providing adequate scalability. In an extensive evaluation based on real-world data from a mobile application, we illustrate that mPSAuth can provide high accuracy with low encryption and communication overhead, while the effort for the inference is increased to a tolerable extent.
\end{abstract}

%% file: sections/introduction.tex
\IEEEraisesectionheading{\section{Introduction}\label{Sec:Introduction}}
\IEEEPARstart{I}{n} today's world, services offered as web applications has become an indispensable part of most people's everyday lives. These include shopping, social media, banking applications, and many more. Most of these applications can be accessed with mobile devices, making them available anytime and almost anywhere. In many cases, the use of these services is closely linked to the digital identity of the user, which must be checked by the provider using a robust authentication process~\cite{ContAuthChallengesN1, ContAuthChallengesN2}. The success of a particular authentication process depends on many characteristics, such as the level of security provided, the protection of the user's privacy, compliance with legal regulations, but also usability and scalability aspects. For these reasons, an advanced authentication process is indispensable for \emph{mobile web applications}. In the context of this work, we define the term mobile web application as an application that meets the following criteria: (1) It is organized as a client-server-based system~\cite{ClientServerBerson1992}. (2) Clients are mobile devices, such as smartphones, whose physical location can change dynamically. (3) Application usage is bound to user identities, which the server must be able to verify.

Currently, the landscape of authentication methods used in mobile web applications is highly diverse. In general, traditional passwords are still the most commonly used method. Unfortunately, the passwords can fall into the hands of an attacker in many ways~\cite{Bang2012PWSecurity}. Starting from this problem, \emph{multi-factor authentication (MFA)} emerged, which requires users to provide multiple pieces of evidence (factors) to confirm their identity~\cite{Ometov2018MF}. However, the main problem associated with MFA is that additional factors, such as sending a message to the user's mobile phone number or mail address, can significantly decrease the application's usability~\cite{Reese2019MFA, Wiefling_Verify_2021}. This problem is exacerbated when considering continuous authentication scenarios~\cite{Shahzad2017ContinuousAuth, ContinuousAuthChallenges2016}, where users are periodically re-authenticated while using the application. One approach to tackle these challenges is \emph{risk-based authentication (RBA)}, which can be seen as a special case of MFA~\cite{WieflingRBA19}. It involves the determination of a risk profile based on information about the user's behavior and device. Depending on the risk profile, the application provider can decide the amount and type of factors the user has to comply with. These are selected to minimize the impact on the application's usability while ensuring an appropriate level of security~\cite{WieflingLongTerm2021}. As it is common for mobile web applications to have high numbers of users, the risk estimation process must scale adequately \textbf{(P1)}. Moreover, behavioral data tends to be highly sensitive, which puts users' privacy at risk~\cite{Wiefling_Privacy_2021} \textbf{(P2)} and requires compliance with data protection laws, such as the GDPR~\cite{Kuner_ML_Law_2017} \textbf{(P3)}.

Several approaches attempt to mitigate these problems in the context of mobile web applications. The most recent and promising of these rely on \emph{machine learning (ML)} techniques to analyze user behavior and thereby provide a basis for authentication decisions, e.g., using data arising on the user's mobile device \emph{(frontend)}~\cite{Karanikiotis2020TouchTraces, acien2019multilock, thao2020gps, TouchTracesPrivacy2013} or the application servers \emph{(backend)}~\cite{Freeman2016WhoAY, Ongun2019AuthIoT}. Popular data sources include sensor data or locations of the mobile device, as well as network addresses and browser information. However, these approaches either do not provide sufficient privacy protection, suffer from scalability issues as the number of ML models to be trained increases linearly with the number of users, or their architecture is not suitable for mobile web applications. Established authentication mechanisms such as facial recognition or fingerprint recognition in their traditional form are also inappropriate for continuous authentication in mobile web applications. These require special hardware and continuous scanning, which can harm usability and privacy~\cite{ContAuthChallengesN2, ContAuthFinger}.

As an approach that addresses these gaps, we introduce \emph{mPSAuth: Privacy-Preserving and Scalable Authentication for Mobile Web-Applications}. It enables continuous RBA for mobile web applications based on a novel authentication protocol that puts user privacy first. The protocol uses homomorphic encryption, which is applied to behavioral data before it leaves the user's device. As a result, strong security guarantees can be given without degrading the accuracy of the authentication. The data sources consulted for gathering behavioral data are generic and can be customized depending on the use case. In addition, the models are reused across all users, ensuring that the number of ML models to be trained is constant concerning the user count. This provides scalability for all types of modern applications.

The primary contributions of mPSAuth are as follows:
\begin{itemize}
    \item[\textbf{(C1)}] Protecting the users privacy by transferring and processing behavioral data only in homomorphically encrypted form (P2). In this way, it is also much easier to meet legal requirements, as the user's personal data does not have to be disclosed (P3).
    \item[\textbf{(C2)}] Reasonable scalability characteristics due to a fixed number of ML models that need to be trained, regardless of the number of users (P1).
    \item[\textbf{(C3)}] Assurance of decent performance in terms of required computational effort and network traffic involved in the authentication protocol.
\end{itemize}

We constructed an extensive evaluation scenario using data collected within a real-world mobile gaming application called \emph{BrainRun}~\cite{brainrun2019}. We evaluated the accuracy of the authentication, demonstrating that the architecture of mPCAuth is suitable for continuous RBA of users in mobile web applications. Subsequently, we investigated the impact of the privacy-preserving authentication protocol computational overhead and network traffic. It turned out that homomorphic encryption causes the inferences to be significantly more time-consuming and the network traffic to increase. Nevertheless, the evaluation results showed that mPSAuth can be applied in practical use cases and still has considerable potential for improving performance in the future, thanks to ongoing development in the field of homomorphic encryption~\cite{FHEFuture1, CheetahSpeedup2021, zhai2021accelerating, RNNHE2020}.

The remainder of this paper is structured as follows. Starting with Section~\ref{Sec:Foundations}, we describe the fundamental concepts and approaches on which mPSAuth relies. Next, Section~\ref{Sec:ThreatModel} describes the threat model and related assumptions that form the basis for security and privacy considerations. Subsequently, Section~\ref{Sec:Approach} presents a detailed description of the architecture of mPSAuth, the underlying authentication protocol, and explains the functionality of the individual building blocks. Section~\ref{Sec:Evaluation:Setup} outlines the structure of the evaluation, followed by Section~\ref{Sec:Evaluation:Results}, which shows and interprets the evaluation results. Thereafter, Section~\ref{Sec:RelatedWork} summarizes related work and compares it to mPSAuth. Finally, Section~\ref{Sec:Conclusion} concludes the central findings and points out future work.

%% file: sections/foundations.tex
\section{Foundations and Background} % Objectives and Background
\label{Sec:Foundations}
In this section, we introduce concepts our work builds on.

\subsection{Homomorphic Encryption}
\label{Sec:Foundations:HomomorphEnc}
Homomorphic encryption~\cite{YiHE2014} is practically an extension of public-key cryptography that allows performing computations on encrypted data. There are various approaches that differ in what operations can be performed on the encrypted data. The schemes that support both additive and multiplicative operations are called fully homomorphic encryption schemes~\cite{Armknecht2015AGT}. These are of great interest since arbitrary operations can be performed, making them viable for many use cases. However, a challenge is the performance of these systems because the operations on the encrypted data are associated with high computational effort~\cite{NaehrigHE2011}. This has prevented the widespread adoption of homomorphic encryption. Nevertheless, homomorphic encryption has regained popularity lately thanks to new homomorphic encryption frameworks such as those offered by IBM~\cite{IBMHomomorph2019} and Microsoft~\cite{sealcrypto}. These are still under active development, which is promising for future improvements, especially in terms of performance.

\subsection{Privacy-Preserving Machine Learning}
\label{Sec:Foundations:PrivacyML}
On top of fully homomorphic encryption schemes, several approaches enable ML model inferences on encrypted data. Typically, the ML model is trained on unencrypted data. Then, the model is transformed to perform the computations on encrypted data, which is possible due to the properties of homomorphic encryption. The inference results are encrypted and can only be decrypted by the entity which possesses the private key. These properties are appealing for use cases where the capabilities of an ML model are offered as a service because the client does not have to fear that his data will be misused.
Both IBM~\cite{IBMHomomorph2019, aharoni2021helayers} and Microsoft~\cite{JowlinCryptoNets2016} offer tools with which ML model inferences can be realized on homomorphically encrypted data. Depending on the tool, neural networks, decision trees, and logistic regressions are supported. However, there are considerable limitations. For example, only specific layers are supported in neural networks, and the complexity of the networks must be kept relatively low to ensure a reasonable throughput.

%% file: sections/threatmodel.tex
\section{Threat Model and Assumptions}
\label{Sec:ThreatModel}
We consider three participating entities in the authentication protocol of mPSAuth:
\begin{itemize}
    \item[(i))] mobile device of user that should be authenticated and which runs the application under observation
    \item[(ii)] backend of the application under observation that wants to authenticate users continuously
    \item[(iii)] application server that manages the authentication
\end{itemize}
The backend (ii) can also coincide with the authentication server (iii), which is a common setting in practice. Nevertheless, we separate the application backend and the authentication server semantically. In this way, business models such as authentication-as-a-service are supported. Consequently, we expect that no trust relationships exist between the mobile device (i), the backend (ii), and the authentication server (iii).

Threats originate from internal adversaries who actively participate in the authentication process as an entity, which means that one or multiple of the three mentioned participants have malicious intentions. All other adversaries are external and are not considered further because they can be handled by conventional network security practices~\cite{TLSTurner2014}.

An adversary's goal can be twofold. On the one hand, he can aim to manipulate the authentication mechanism to impersonate a legitimate user (\textit{A1}). However, this is only reasonable from the point of view of the mobile device (i) that wants to authenticate itself. Because the backend (ii), as well as the authentication server (iii), intend to authenticate users correctly. Here, we do not address takeover scenarios of the authentication server and the backend, as the authentication process could be changed at will. This would allow an attacker to modify the authentication decision as required. On the other hand, the goal of a curious entity may be to obtain sensitive data by applying the authentication protocol (\textit{A2}). All participating entities may be curious. For example, a curious backend (ii) or authentication server (iii) may be interested in using sensitive information for user profiling. Similarly, a mobile client device may be interested in mining behavioral data of other users.

The capabilities we consider for adversaries to achieve their respective goals are as follows. Both attacker types (A1, A2) will eventually not conform to the specified protocol (malicious adversary~\cite{Do2018TheRO}). This means that they can behave actively and modify, add, or remove protocol messages. Logically, the attacker can change the application's code on his device to achieve his goal. Of course, both attacker types can also behave passively by carrying out the protocol honestly, with the goal to bypass the authentication (A1) or to gather sensitive details from the exchanged messages (A2). Moreover, we include attackers who take over a legitimate user's application session, e.g., by session hijacking or stealing a device. However, we assume that the attacker does not have detailed information about the legitimate user's behavior before the takeover and does not obtain such information as a result. Otherwise, A2 would already be successful by implication, as he obtained behavioral information from one or more users. For A1, we explicitly exclude attacks that involve substantial knowledge about the behavior of other users (e.g., replay attacks). Actually, other biometric authentication methods, including face recognition and fingerprint recognition, are also vulnerable to this kind of attack~\cite{FaceIDAttacks2021, FingerprintAttacks2021}.

We analyzed the characteristics of our approach in Section \ref{Sec:PrivacyProt} based on the threats and associated assumptions described in this section to show that potential adversaries can be mitigated effectively.

%% file: sections/approach/approach.tex
\section{The mPSAuth Approach}
\label{Sec:Approach}
In this section, we present the architecture of our approach, with descriptions of the underlying components.

%% file: sections/approach/overview.tex
\subsection{Approach Overview}
\label{Sec:Overview}
% short intro
The mPSAuth approach is intended to provide continuous and risk-based authentication of users within mobile web applications. For this purpose, the user's behavior within the application is regularly investigated. We analyze various data sources that reflect user behavior either separately or jointly using machine learning techniques. To this end, we compare whether the currently observed behavior matches the known behavior of the user. We then base our authentication decision on the result of this comparison.

% enrollment part
Because our approach requires information about the user's previous behavior, an enrollment phase is necessary. During this phase, conventional authentication factors such as passwords combined with SMS verification must be used. Following the enrollment phase, the ML-based analysis of behavioral data can serve as a stand-alone authentication factor. Sporadic addition of further authentication factors (e.g., SMS verification) is intended to cover exceptional cases. For example, in the case that a legitimate user is mistakenly rejected repeatedly due to a significantly changed behavior. The duration of the enrollment phase depends on the extent of the baseline to be collected (see Section~\ref{Sec:Detectors}).

A key feature of our approach is scalability, as mPSAuth trains the required models so that they can be applied for all users. Consequently, the effort needed to establish the models is constant for an increasing number of users. Since behavioral data contains highly sensitive information, mPSAuth uses a privacy-preserving authentication protocol built on fully homomorphic encryption. This implies that the inference of the ML models used must be able to work directly on homomorphically encrypted data.

\begin{figure*}[!htb]
	\centering
  \includegraphics[width=1\textwidth]{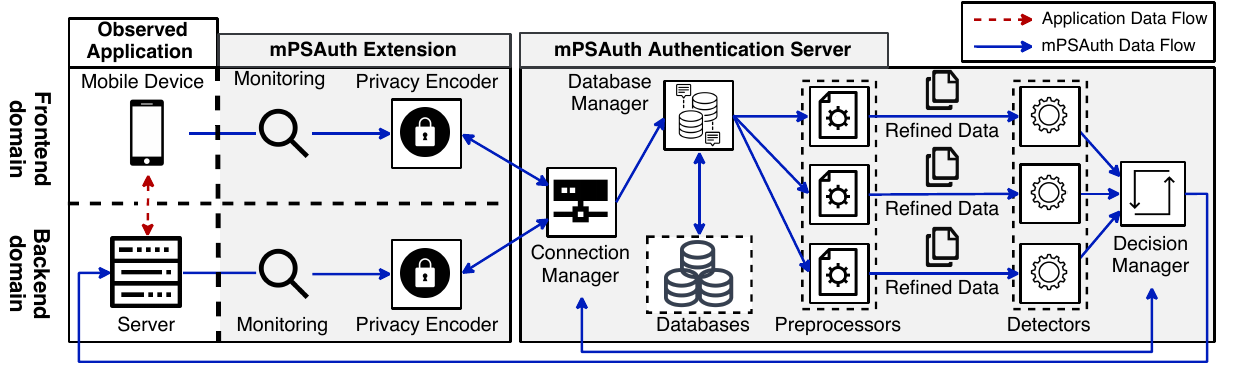}
	\caption{Overview of the main components and data flows that are part of the mPSAuth architecture}
	\label{Fig:Conception:Overview}
\end{figure*}

Figure~\ref{Fig:Conception:Overview} visualizes the architecture of mPSAuth, including the data flows between the key components. Starting on the left-hand side, we consider the ecosystem of the mobile web application in the context in which mPSAuth is applied. We distinguish between the frontend that provides the application to the user (mobile device) and the backend that provides the required services (server). In order to authenticate users according to their behavior, it is necessary to collect appropriate data sources. The collection of the behavioral data is realized by means of a \emph{monitoring} that is integrated into the parts of the application that run in the frontend and backend. The data sources we consult are introduced in Section~\ref{Sec:DataCollection}. 

Additionally, mPSAuth introduces the \emph{privacy encoder} for integration into the frontend and backend. It is the central element that ensures the privacy of the collected data. This is achieved by homomorphically encrypting the behavioral data that leaves the respective domain. Thus, the \emph{authentication server} can perform calculations on it but cannot draw any conclusions about the underlying sensitive data. Section~\ref{Sec:PrivacyProt} presents details on how the data is encrypted and can still be used for authentication.

% new block beginning with authentication server stuff
Subsequently, the privacy encoders transmit the encrypted data to the \emph{connection manager}, which resides on the authentication server. The connection manager passes the received behavioral data of the users to the \emph{database manager}, which is responsible for storing it in a well-structured way. Furthermore, the database manager is accountable for providing the data in a suitable format for the \emph{preprocessors}.

Each of the preprocessors prepares the data necessary to investigate one aspect of the user's behavior. mPSAuth executes the authentication at regular intervals and/or when the user performs a specific action that should be explicitly authenticated (e.g., purchase completion). When an authentication is triggered, the preprocessors fetch the most recently collected data points (\emph{observation}) and a baseline of data points (\emph{history}) from the database manager. The combination of history and observation, called \emph{refined data}, is forwarded to the \emph{detectors}, whose task is to check if they are consistent. For checking the consistency of the observation with the history, the detectors rely on ML methods. The conceptual design of the input for the detectors enables a single model to be trained per detector, which can be used for all users, thus ensuring scalability for a high number of users. However, the refined data is only available in homomorphically encrypted form. Therefore, the ML method used must be able to perform the inference on encrypted data. It also follows that the inference result is encrypted, which is why mPSAuth decrypts it by an exchange between the connection manager and the respective privacy encoder. For this purpose, \emph{verifiable decryption}~\cite{Fucai2018VerDec} is used to ensure that the connection manager receives the correct value. Section~\ref{Sec:PrivacyProt} describes this procedure and the methodology for protecting privacy more closely.

The decryption of the final result is coordinated by the \emph{decision manager}, which also aggregates the values obtained from the individual detectors. The goal is to determine an overall risk level that reflects the likelihood of the current user being illegitimate. Finally, the risk level is sent to the application server as input for an authentication decision. The interactions between the preprocessors, the detectors, and the decision manager are described in Section~\ref{Sec:Detectors}.

%% file: sections/approach/monitoring.tex
\subsection{Behavioral Data Collection}
\label{Sec:DataCollection}
We consider data sources from both the frontend and the backend of the application to obtain information about the users' behavior. The conceptual design allows data sources to be added or removed easily, resulting in a high degree of flexibility regarding the types of devices to which mPSAuth can be applied. The following list summarizes relevant data sources and explains their general structure. The selection builds on existing work that has investigated and ensured the suitability for behavior-based authentication.

\begin{itemize}[leftmargin=*]
    \item \textbf{Frontend Data Sources:}
    \begin{itemize}[leftmargin=*]
        \item \textit{Touchscreen Data}: Related work shows that the patterns observed when the user interacts with the touchscreen are highly individual, and therefore can be used for user identification~\cite{Karanikiotis2020TouchTraces, acien2019multilock}. For this reason, we capture the general properties of the touchscreen (e.g., size and pixel density) and the points at which the user performs taps or swipes. Additionally, continuous recording allows us to derive features such as the movement speed.
        \item \textit{Sensor Data}: Sensors are an essential part of mobile devices to enable intuitive handling. They also reflect how users interact with their devices and can serve as an authentication factor~\cite{mci/Li2018, acien2019multilock, acien2020smartphone}. Motivated by this, we consider common sensors such as accelerometers, magnetometers, and gyroscopes as data sources.
        \item \textit{Location and Network Connection Data}: In contrast to conventional computers, mobile devices usually change their location frequently, and the resulting movement profiles are strongly dependent on the specific user. Hence, both conventional~\cite{Denning1996LocationbasedAG, Zhang2012LocAuth} and novel authentication systems~\cite{thao2020gps, acien2019multilock} rely on the physical locations for authentication purposes. User locations can be determined in several ways, the most common being GPS~\cite{GPS2002}, but information about the current network connection can also approximate them~\cite{tennekes2021bayesian}.
    \end{itemize}
    \item \textbf{Backend Data Sources:}
    \begin{itemize}[leftmargin=*]
        \item \textit{Network Traffic}: From the backend's perspective, traffic received at all layers of the network stack can indicate a user's identity. For example, HTTP request parameters such as the user agent can reveal information about the user's device~\cite{UserAgents2012}, or data from lower layers, such as network addresses, can disclose the user's location~\cite{Freeman2016WhoAY}.
    \end{itemize}
\end{itemize}

The list is non-exhaustive because the architecture of mPSAuth is designed to be flexible so that other data sources can be incorporated. Therefore, it would also be possible to include traditional biometric data sources, such as fingerprints or facial data. Ultimately, a sound authentication decision requires consideration of several data sources.

%% file: sections/approach/privacy.tex
\newenvironment{conditions*}
  {\par\vspace{\abovedisplayskip}\noindent
   \tabularx{\columnwidth}{>{$}l<{$} @{\ : } >{\raggedright\arraybackslash}X}}
  {\endtabularx\par\vspace{\belowdisplayskip}}

\subsection{Privacy-Preserving Authentication}
\label{Sec:PrivacyProt}
The core of mPSAuth is an authentication protocol that allows the authentication server to analyze the user's behavioral data collected in the frontend and backend without exposing sensitive information. The adoption of homomorphic encryption can guarantee this feature, thus increasing the user's trust in the authentication procedure~\cite{Jones07PercAuth}.

\subsubsection{Authentication Protocol}
The prerequisite for our authentication protocol is a fully homomorphic encryption scheme, where two different keys are used for encryption and decryption (see Section~\ref{Sec:Foundations:HomomorphEnc}).

\begin{figure}[!htb]
	\centering
  \includegraphics[width=1\linewidth]{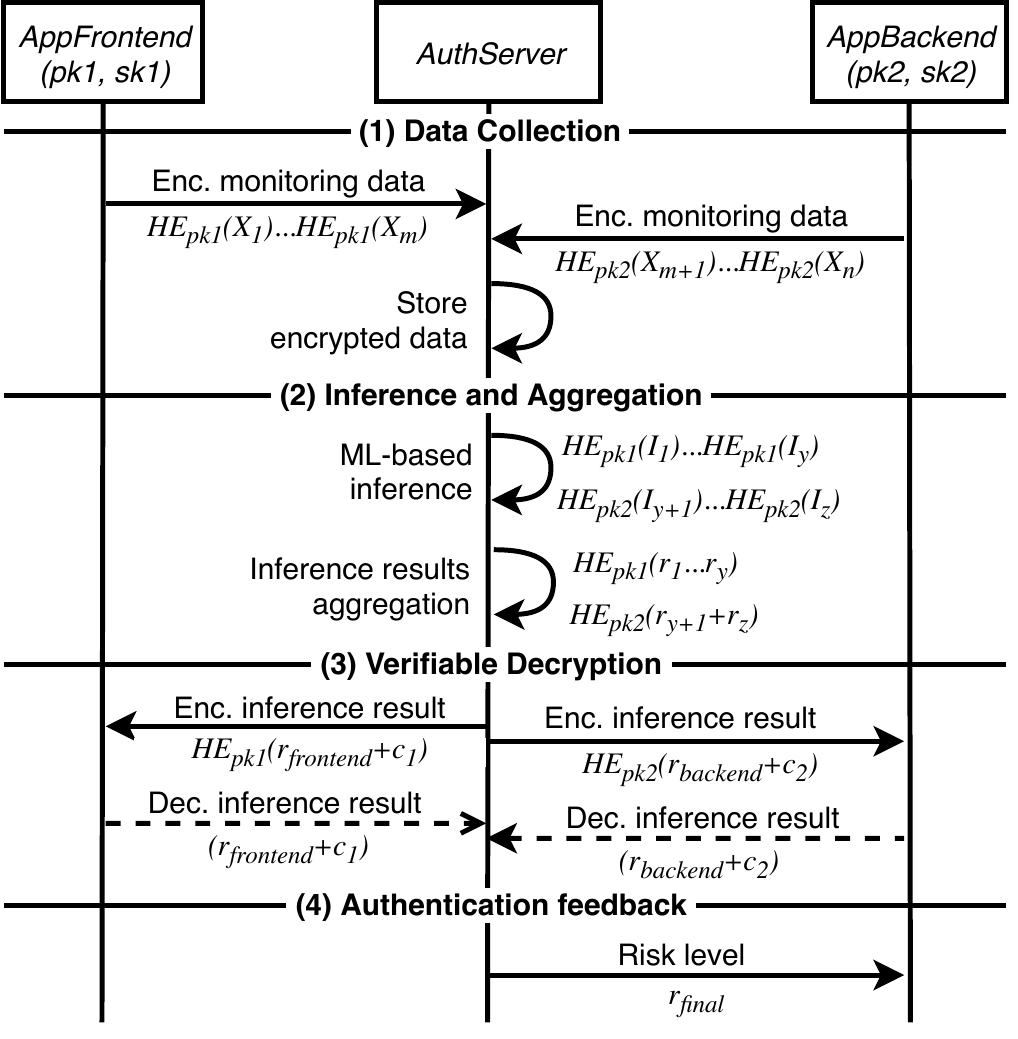}
	\caption{Message sequences involved in our privacy-preserving authentication protocol}
	\label{Fig:Conception:Privacy:Sequence}
\end{figure}

Basically, we distinguish between four phases in the protocol: first, the phase in which behavioral data is collected, followed by three phases that are carried out when the risk-based authentication is triggered. mPSAuth continuously performs the data collection, while the authentication process is invoked at regular intervals or on explicitly defined events. \autoref{Fig:Conception:Privacy:Sequence} summarizes the messages exchanged between the entities within the phases (cf. threat model from Section~\ref{Sec:ThreatModel}). In this context, we denote data $D$ that has been homomorphically encrypted with a key $k$ by $HE_{k}(D)$. We assume that the user possesses a key pair consisting of a public key ($pk1$) and a private key ($sk1$). In addition, $pk1$ must be communicated to the authentication server once, e.g., within the registration process. The same applies to the backend, which also has its own key pair ($pk2, sk2$).

In the phase of collecting behavioral data (1), data sources introduced in Section~\ref{Sec:DataCollection} are monitored, and the results ($X_{1} \dots X_{n}$) are homomorphically encrypted using the respective public keys ($HE_{pk1}(X_{1})$ $\dots$ $HE_{pk2}(X_{n})$). Every $X_{i}$ is a matrix reflecting the data collected from one data source, whose rows contain one data point each, and the columns store the related features. Next, the encrypted data ($HE_{pk1}(X_{1}) \dots HE_{pk2}(X_{n})$) is transmitted to the authentication server, which stores it.

Once an authentication process is triggered (2), the authentication server retrieves the encrypted behavioral data from the database and transforms it into matrices ($I_{1} \dots I_{z}$), each of which encapsulates behavioral data to be analyzed. It should be noted that the number of derived matrices (z) may well exceed the number of existing data sources. This is covered in the description of the analysis procedure in Section~\ref{Sec:Detectors}. Next, the matrices ($I_{1} \dots I_{z}$) are used as input to ML models that are capable of operating on the homomorphically encrypted data and aim to determine whether the currently observed behavior conforms to the known behavior of the user. We elaborate on the conceptual design of these models in Section~\ref{Sec:Detectors}. Thereafter, the authentication server aggregates the encrypted inference results ($HE_{pk1}(r_{1}) \dots HE_{pk2}(r_{z})$), using an arbitrary function that is supported by the encryption scheme (see Section~\ref{Sec:Detectors}). This is done separately for the inference results obtained from frontend data ($HE_{pk1}(r_{1}) \dots HE_{pk1}(r_{y})$) and backend data ($HE_{pk2}(r_{y+1}) \dots HE_{pk2}(r_{z})$), as these are encrypted with different keys. The outcome of the aggregations are two values ($HE_{pk1}(r_{frontend})$ and $HE_{pk2}(r_{backend})$), depending on the origin of the underlying data.

The aggregated results are homomorphically encrypted ($HE_{pk1}(r_{frontend})$ and $HE_{pk2}(r_{backend})$) and cannot be decrypted by the authentication server because it lacks the private key of both the frontend and backend. For this reason, the authentication protocol of mPSAuth requires an interaction with the frontend and the backend (3). Initially, the authentication server adds random numbers ($c_{1}, c_{2}$) to $HE_{pk1}(r_{frontend})$ and $HE_{pk2}(r_{backend})$, which is possible due to the properties of homomorphic encryption. The purpose of the random numbers is explained in the security analysis. Then, the modified results ($HE_{pk1}(r_{frontend} + c_{1})$ and $HE_{pk2}(r_{backend} + c_{2})$) are sent back to the frontend or the backend, respectively. These decrypt the results ($(r_{frontend}+c_{1}), (r_{backend}+c_{2})$) and send them back to the authentication server. To prevent the decrypting party from manipulating the authentication by returning modified values, we apply an efficient verifiable decryption based on a zero-knowledge proof presented by Luo et. al.~\cite{Fucai2018VerDec}. Finally, the authentication server subtracts the randomly added values and aggregates the decrypted inference results ($r_{frontend}, r_{backend}$) to calculate a final risk level ($r_{final}$), which is then transmitted to the application backend.

\subsubsection{Security Analysis}
In the following, we analyze the security characteristics of the presented authentication protocol with respect to the threat model defined in Section~\ref{Sec:ThreatModel}. Thereby, we substantiate \emph{contribution 1 (C1)} of mPSAuth.

First, we consider the situation when the frontend is controlled by an attacker who tries to impersonate a user (\textit{A1}). As stated in the threat model, this objective is not relevant for the other entities, as they are concerned with correct authentication. The attacker can act either passively or actively. As a passive attacker must comply with the protocol, the only option is to imitate the behavior of the legitimate user. As we assume that the behavior is unknown to the attacker, he can either generate random data or keep his behavior unmodified. Both cases are mitigated by analyzing the behavior with ML techniques as described in Section~\ref{Sec:Detectors}. We show the accuracy of the detection within our evaluation (see Section~\ref{Sec:Evaluation:Results:Accuracy}).

An active attacker can perform non-honest decryption of the aggregated inference result ($HE_{pk1}(r_{frontend}+c_{1})$) requested by the authentication server. In other words, he sends a forged inference result to the authentication server to distort the risk level calculation. In the proposed protocol, this is prevented by verifiable decryption~\cite{Fucai2018VerDec}. It ensures that the frontend proves to the authentication server that the value sent is the correct decryption ($r_{frontend}+c_{1}$) of the encrypted inference result.

Second, we analyze the attack scenario where a participating entity wants to obtain sensitive data by applying the protocol (\textit{A2}). Initially, we consider the perspective of the frontend. It is relevant whether an attacker can use the authentication protocol to gain possession of behavioral data from other users. The aggregated inference result ($HE_{pk1}(r_{frontend}+c_{1})$) comprises the only information transferred from the authentication server to the frontend. At this point, it must be distinguished whether the attacker knows the private key of the user under attack. If he does not know the private key, he cannot decrypt the inference results and thus gain any information. If he knows the private key (e.g., by compromising the user's end device), he can interpret the inference result ($r_{frontend}+c_{1}$) to determine whether the behavioral data he sent to the authentication server is consistent with that of the user. However, this is mitigated by adding a random value ($c_{1}$) to avoid that the attacker can interpret the results. Logically, the only information an attacker obtains is whether or not authentication was successful. Due to the high dimensionality of the behavioral data (cf. Section~\ref{Sec:Evaluation:Setup:Experiment}), bruteforce attacks that exploit this are not efficient and can be easily mitigated by blocking authentication requests after a certain number of failed attempts. The same insights apply when considering \textit{A2} from the backend perspective.

Finally, we analyze \textit{A2} from the perspective of a malicious authentication server that aims to collect sensitive user behavior data. The server explicitly receives behavioral data from the frontend and the backend in encrypted form ($HE_{pk1}(X_{1})$ $\dots$ $HE_{pk2}(X_{m})$). Consequently, the protection of this data depends on the homomorphic encryption scheme applied. In general, if the authentication server does not know the private key of the user or the backend, it cannot infer the content of the encrypted data (see Section~\ref{Sec:Foundations:HomomorphEnc}). 

An alternative attack is to modify the protocol and send encrypted behavioral data instead of the encrypted inference results ($HE_{pk1}(r_{frontend})$,$HE_{pk2}((r_{backend})$). As these are decrypted and sent back, the authentication server can obtain parts of behavioral data. However, the number of values for which the authentication server can request decryption is limited to a fraction of the amount of behavioral data sent. The experiment within our evaluation (see Section~\ref{Sec:Evaluation:Setup}) confirms that the amount of data that can be decrypted by the authentication server is far smaller than the amount of behavioral data involved in an authentication process. By implication, the authentication server needs a lot of steps to learn anything meaningful about the users' behavior. Moreover, such a change to the protocol would cause the authentication server to lose its ability to calculate a proper risk level, as it can not interpret the inference results without communicating them to the other parties.

%% file: sections/approach/detectors.tex
\subsection{Scalable ML-based Behavior Analysis}
\label{Sec:Detectors}
In this section, we review the mechanisms that enable the use of machine learning techniques on homomorphically encrypted data to check whether a user's currently observed behavior matches their past behavior. These are fundamental to the authentication protocol introduced in the previous Section~\ref{Sec:PrivacyProt}. We also highlight how mPSAuth keeps the number of trained ML models constant regardless of user count, thereby ensuring appropriate scalability properties.

\subsubsection{Data Preprocessing}
\begin{figure}[!htb]
	\centering
  \includegraphics[width=0.95\linewidth]{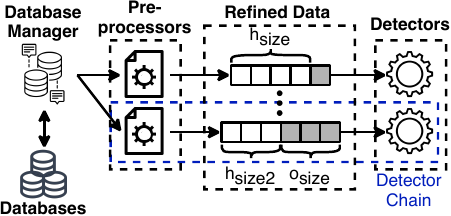}
	\caption{Preparation of the behavioral data in order to analyze them on the basis of ML techniques}
	\label{Fig:Conception:ML:DetectorChain}
\end{figure}

The preprocessors are responsible for converting the collected data about a user into a structured form (\emph{refined data}), which can be used as input for the detectors (see Figure~\ref{Fig:Conception:Overview}). Every detector is linked to exactly one preprocessor, which requests a number of data points observed in previous sessions (\textit{history window}) and a number of data points observed in the current session (\textit{observation window}). We refer to the totality of history and observation as \textit{investigation window}. Figure~\ref{Fig:Conception:ML:DetectorChain} illustrates this procedure.

The number of elements in the history window ($h_{size}$) and the observation window ($o_{size}$) are parameters that may depend on the aspect of user behavior that should be analyzed. The combination of a preprocessor and a detector is called a \textit{detector chain} and depending on the amount of data that needs to be examined for consistency with the known behavior of a user, it can be triggered multiple times.

\subsubsection{ML-based Behavior Analysis}
The detectors receive an investigation window as input and are supposed to decide whether the user's current behavior is anomalous based on the history as a baseline. The formal representation of the input for the single execution of a detector looks as follows:
\[
\begin{array}{l}
IW =
\left[\begin{smallmatrix}HE_{k}(h_{1}) & \dots & HE_{k}(h_{a}) \end{smallmatrix}\middle|
\begin{smallmatrix}HE_{k}(o_{1}) & \dots & HE_{k}(o_{b})\end{smallmatrix}\right]
\end{array}
\]
In this context, $IW$ represents the investigation window, $h_{1} \dots h_{a}$ and $o_{1} \dots o_{b}$ describe the data points within the history window and the observation window, respectively. Furthermore, $a$ is the size of the history window ($a = h_{size}$), and $b$ is the size of the observation window ($b = o_{size}$). The execution of the detector leads to a homomorphically encrypted inference result ($HE_{k}(r_{i}))$), which is incorporated into the risk level (see Section~\ref{Sec:PrivacyProt}). mPSAuth relies on ML-based detectors capable of operating on homomorphically encrypted data as presented in Section~\ref{Sec:Foundations:PrivacyML}. The method to be used depends on the type of data to be analyzed and the considered use case. For example, some methods may perform different depending on the behavioral data (e.g., sensor data) or application type (e.g., mobile games).

An advantage of mPSAuth is that the trained models can be applied across all users. In other words, only one model needs to be trained per detector, compared to related approaches that train a custom model to detect the behavior of a particular user. Our strategy incorporates the user's behavioral history into the input of the model, whereas other approaches incorporate it into a model throughout the training process. Although this may result in an accuracy loss, it is compensated by combining multiple detectors and an enormous scalability advantage in terms of training effort. Recall that due to homomorphic encryption training of the models must be carried out on the raw data (see Section~\ref{Sec:Foundations:PrivacyML}). Our models can first be trained with publicly available datasets and then applied in production for all users for privacy-compliant inference. This is a huge advantage over approaches that train one model per user, which would require raw data from each user that cannot be obtained without violating privacy.

\subsubsection{Risk Level Estimation}
The final step towards a summarized risk level is to merge the encrypted inference results obtained from the executions of the detectors ($HE_{k}(r_{1}) \dots HE_{k}(r_{z})$). The risk level estimation is performed by the decision manager, which was introduced in \autoref{Fig:Conception:Overview}. It receives an arbitrary number of inference results from each detector, depending on the amount of data and the size of the observation window. It follows that the method used to aggregate the inference results must handle varying input sizes and needs to be supported by the encryption scheme. Possible implementations range from simple techniques, such as calculating means, to more complex ones, such as ML models. Therefore, mPSAuth considers the method applied to be interchangeable (strategy pattern). In the end, the calculated risk levels are reported to the application server (cf. Section~\ref{Sec:PrivacyProt}).

%% file: figures/evaluation/values.tex
% Input Dimensions
\newcommand{\MinInputSize}{2535}
\newcommand{\MaxInputSize}{11700}

% Authentication Accuracy
\newcommand{\AccuracyMaxGyroscope}{29.70\%}
\newcommand{\AccuracyMinGyroscope}{20.35\%}

\newcommand{\AccuracyMaxAccelerometer}{15.22\%}
\newcommand{\AccuracyMinAccelerometer}{11.24\%}

\newcommand{\AccuracyMaxMagnetometer}{33.46\%}
\newcommand{\AccuracyMinMagnetometer}{27.40\%}

\newcommand{\AccuracyMaxBackend}{20.55\%}
\newcommand{\AccuracyMinBackend}{13.45\%}

\newcommand{\AccuracyMaxSwipes}{30.91\%}
\newcommand{\AccuracyMinSwipes}{24.71\%}

% Monitoring Data Overhead
\newcommand{\EncryptionDataOverheadMin}{0.625}
\newcommand{\EncryptionDataOverheadAvg}{7.935}
\newcommand{\EncryptionDataOverheadMax}{73.787}

\newcommand{\RawDataOverheadMin}{0.004}
\newcommand{\RawDataOverheadAvg}{0.381}
\newcommand{\RawDataOverheadMax}{3.674}

\newcommand{\EncryptionDataOverheadRatio}{20.83}

% Inference Time
\newcommand{\InferenceTimeRaw}{0.24 seconds}
\newcommand{\InferenceTimeEnc}{5.16 seconds}
\newcommand{\InferenceTimeRatio}{21.5}

%% file: sections/evaluation/setup.tex
\section{Evaluation Setup}
\label{Sec:Evaluation:Setup}
We apply the Goal Question Metric (GQM)~\cite{Basili94} approach to investigate the accuracy and performance of mPSAuth.

\subsection{Evaluation Goals and Research Questions}
\label{Sec:Evaluation:Setup:Goals}
We focus on two key aspects of our approach. First, we evaluate the accuracy of the risk-based authentication on behavioral data collected by mobile devices (\textbf{Goal 1}). This feature is fundamental for ensuring the security of an application using the authentication system. Second, we analyze the performance characteristics of mPSAuth in a realistic scenario, as they are crucial in real-world applications (\textbf{Goal 2}). Accordingly, the study of these two goals is broken down into the following research questions (RQs):
\begin{itemize}[leftmargin=*]
    \item \textbf{Goal 1: Risk-based authentication accuracy:}
    \begin{itemize}
        \item \textbf{RQ-1.1}: How well do different data sources perform as bases for authentication with a single detector?
        \item \textbf{RQ-1.2}: How accurate is mPSAuth in a continuous authentication setting when combining detectors operating on different data sources?
    \end{itemize}
    
    \item \textbf{Goal 2: Risk-based authentication performance:}
    \begin{itemize}
        \item \textbf{RQ-2.1}: What is the computational effort for encrypting the monitoring data homomorphically?
        \item \textbf{RQ-2.2}: How does the authentication protocol affect the communication effort between the parties involved?
        \item \textbf{RQ-2.3}: How significant is the imposed performance overhead associated with ML model inference on homomorphically encrypted data?
    \end{itemize}
\end{itemize}

% Block Goal 1
RQ-1.1 intends to evaluate whether the data sources used are generally suitable for behavior-based authentication. Together with RQ-1.2, which investigates the combination of different data sources and their applicability for continuous user authentication, it is shown that mPSAuth is a promising approach for risk-based and continuous user authentication in mobile web applications (\emph{contribution 1}).

% Block Goal 2
The three research questions RQ-2.1, RQ-2.2, and RQ-2.3, address the performance and scalability characteristics of mPSAuth (\emph{contribution~2}, \emph{contribution~3}). These are essential with regard to practical use because disproportionately high computational effort is reflected in the financial costs incurred. Beyond that, it must be ensured that the performance of mobile devices, which is generally limited in terms of computational power and network speed, is sufficient.

\subsection{Evaluation Metrics}
\label{Sec:Evaluation:Setup:Metrics}
All research questions that belong to Goal 1 can be reduced to a classification problem, namely whether a user is successfully authenticated or not. For comparability with related approaches, we base the assessment of the research questions on the \emph{Equal Error Rate (EER)}, which is widely accepted in the context of authentication systems~\cite{TEH2016210}. It is defined as the value of the \emph{false acceptance rate (FAR)} and the \emph{false rejection rate (FRR)} when the acceptance threshold is set to a value where both are equal. Here, the FAR describes the percentage of authentications where an illegitimate user is accepted, whereas the FRR describes the percentage of authentications where a legitimate user is rejected.

The answers to research questions RQ-2.1 and RQ-2.3 are based on time measurements that reflect the computational effort required on different platforms. For RQ-2.2, we monitor the absolute amount of data that needs to be transferred when applying the authentication protocol.

\subsection{Evaluation Experiment}
\label{Sec:Evaluation:Setup:Experiment}
We designed an extensive experiment using data from a real-world application. Homomorphic encryption and inference on encrypted data uses Microsoft SEAL~\cite{sealcrypto}.

\subsubsection{Dataset Overview}
The data we use was collected in a mobile gaming application called \enquote{BrainRun}~\cite{brainrun2019} and has also been used by related work to address similar questions. The goal of BrainRun is to create a database to drive research on authentication methods using behavioral data by recording interactions of real users with a mobile game. Two data sources were observed closely: touchscreen gestures (taps, swipes) and sensor data (e.g., accelerometer, magnetometer). In addition, metadata such as the type of device used (e.g., operating system, display size) and personal information about the user (gender, age) were collected. Moreover, the data collection did not take place in a controlled environment, rather the application was publicly available, and data was collected from 2218 different individuals. Therefore, the findings of the evaluation based on this dataset are more meaningful, as the scenario considered is close to reality.

% 2. Dataset foundations
\subsubsection{Dataset Preparation}
In our experiment, we treat the data as if it was obtained by the monitoring of mPSAuth. This allows us to simulate authentication scenarios using our proposed architecture. The data sources we utilize on the frontend side are \emph{swipe gestures} performed by users and data from three different sensors: \emph{accelerometer}, \emph{gyroscope}, and \emph{magnetometer}. Unfortunately, the dataset does not contain any information collected in the backend, and there is a general lack of publicly available data sets in this area. Therefore, we synthesized a minimal HTTP request log from the existing data. In doing so, we proceeded as follows:
\begin{enumerate}[leftmargin=*]
    \item For each of the user's touchscreen actions, we extracted the name of the view on which they were performed. We mapped this information to a fictional HTTP page address, and each time the user navigates to a new view, a HTTP request was created. For example, if the user navigates to the view named \textit{HomeScreen}, an associated HTTP request is created with the page address \enquote{/home}.
    \item Based on the metadata about the devices, we added a user agent to the HTTP requests. In our case, however, it consists only of the operating system.
    \item We added a request parameter that indicates whether the user has finished a game, and the score he achieved.
\end{enumerate}
This gives us five data sources, four of which are captured in the frontend, and one is artificially generated to represent the backend view. Furthermore, we removed users from the dataset who had very few interactions with the application and could not be reasonably included in the experiment.

\subsubsection{Authentication Setting}
% 3. Scenario and Data Preparation
We assume a continuous authentication environment where the identity of the users is regularly validated while they interact with the application. Consequently, we divide the behavioral data of each user into sessions (given by the dataset) and further subdivide the sessions into fractions of 5 minutes (referred to as slices). Every 5 minutes, the data accumulated in the respective slice is used to make an authentication decision. Initially, the data points of all data sources are assigned to a slice, depending on the time they were recorded. We also aggregated the sensor samples into chunks of one second, i.e., ten elements per chunk, as the sampling rate is 100 milliseconds (ms)~\cite{brainrun2019}.

\subsubsection{Data Augmentation}
\begin{figure}[!htb]
	\centering
  \includegraphics[width=0.93\linewidth]{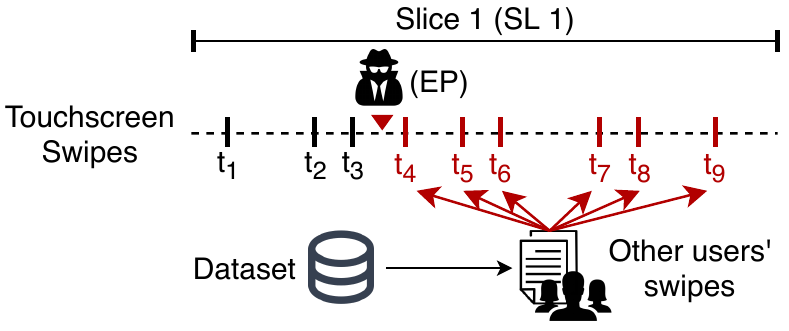}
	\caption{Artificial generation of attack scenarios}
	\label{Fig:Evaluation:Setup:Augmentation}
\end{figure}

The BrainRun dataset only reflects actions performed by legitimate users and does not contain any labeled data representing an attack scenario where a user was impersonated. For this reason, we applied the "one-vs-the-universe" strategy~\cite{Karanikiotis2020TouchTraces} and artificially created slices where the actual user behavior was replaced by that of other users. Figure~\ref{Fig:Evaluation:Setup:Augmentation} shows an example of this procedure for a given slice and the touchscreen swipes as data source.

Note that the data elements used as a substitute must be a consecutive sequence from a single user, as otherwise inconsistencies in the behavior within a slice could mislead the training process of an ML model. We replace a user's behavioral data with that of another user in a slice starting from an \emph{entry point~(EP)}. We vary the EP in each slice and report the respective results separately (see Section~\ref{Sec:Evaluation:Results}).

\subsubsection{Feature Extraction}
For the data sources to be usable as input for ML procedures, suitable features must be selected and extracted. These should be informative, avoid redundancies, and be dimensionally reduced compared to the raw data~\cite{KhalidFeatureExtr2014}. The feature sets used for the data sources are:
\begin{itemize}[leftmargin=*]
    \item \textbf{Touchscreen Swipes:} As there exists profound related work in this area, our feature selection is based on the work of Karanikiotis et al.~\cite{Karanikiotis2020TouchTraces}. Among other features, it recommends the average acceleration, the length, and the deviations from a straight line of the finger movement during a swipe~\cite{Karanikiotis2020TouchTraces}. Additionally, we have extended these features by the area covered by the swipe and its curvature.
    \item \textbf{Accelerometer, Magnetometer and Gyroscope:} The feature extraction is the same for all types of sensor data because the raw samples are equivalently structured and contain the numerical values for the X, Y, and Z direction, respectively. We included these in unchanged form as features to be able to assess the movements precisely.
    \item \textbf{Backend Requests:} Regarding the backend data, we applied one-hot encoding~\cite{Potdar2017ACS} to the requested page address and the operating system contained in the user agent. The request parameters were kept unchanged as features.
\end{itemize}

After extracting the features from the raw data, these were scaled appropriately to speed up the training of the ML models and obtain models with higher fitness~\cite{ZhengScaling2018}.

\subsubsection{ML-based Authentication}
% 4. Models and Preprocessors
In the next step, the detector chains of mPSAuth must be established, i.e., the combination of preprocessors and detectors as presented in Section~\ref{Sec:Detectors}. We analyze all five available data sources separately with one detector chain per data source. Therefore, the combination of the different data sources takes place in the last step when aggregating the results returned by the detectors.

To prepare the data for the detectors, the preprocessors retrieve a \textit{history window} of data points from the data source under consideration and the data points of the respective slice (\textit{observation window}). When constructing the history window, we ensure that the data points are taken from previous sessions. Depending on the observation sizes used for the data sources, the data points in the slice are converted into one or more inputs for the corresponding detector.

The ML models used within the detectors to analyze the behavioral data are structured identically for all five data sources. Due to the limitations of current frameworks for ML inferences on homomorphically encrypted data, we chose to use convolutional neural networks (CNNs)~\cite{MunirCNNAnomaly2019}. We use two convolutional layers with a consecutive average pooling; which are followed by a fully connected layer and a final one-dimensional output layer. We performed the training of the models on 70\% of the users (722) and the measurement of the evaluation results on the other 30\% (310). The evaluation results are highly meaningful as they demonstrate that our models work across users, even if the user's data was not seen in the training phase.

Lastly, the results of the individual detectors have to be bundled into a single risk level. There can be an arbitrary number of results per detector, depending on how many inferences have to be performed, which in turn depends on the encountered elements in the slice and the chosen observation size. We used a rudimentary function to merge the results: first, we calculate the mean of the outputs of each detector and then form the average of the mean values. The aggregated risk level can then be used as a basis for an authentication decision for the respective slice.

%% file: sections/evaluation/results.tex
\section{Evaluation Results}
\label{Sec:Evaluation:Results}
In this section, we present and interpret the results obtained by means of the experimental setup described above.

\subsection{Authentication Accuracy}
\label{Sec:Evaluation:Results:Accuracy}
First, we examined the accuracy of single model inferences depending on the data source used. We varied the history ($h_{size}$) size and the observation size ($o_{size}$) to see how these affect the accuracy. To do so, we trained one model per combination of the sizes of the history window and the observation window and subsequently evaluated it. As a dataset, we constructed the investigation windows of all users and converted 50\% of them into inputs representing an illegitimate user using the procedure described in Section~\ref{Sec:Evaluation:Setup:Experiment}. The size of the composite input for the models ranged from \MinInputSize{} to \MaxInputSize{} in our experiment, depending on the sizes of the history window and the observation window. The minimal input size of \MinInputSize{} emphasizes that the dimension of the behavioral data is sufficiently large that the authentication server cannot gain any substantial information (cf. Section~\ref{Sec:PrivacyProt} - Security Analysis). Accordingly, a malicious authentication server could decrypt at most $\frac{1}{\MinInputSize{}} = 0.039\%$ of the received behavioral data, which we consider to be not significant.

\input{figures/evaluation/accuracy/eer_datasources_reduced_split}

\autoref{Fig:Evaluation:Accuracy:Single} shows the EERs for gyroscope (\textbf{\AccuracyMinGyroscope{}-\AccuracyMaxGyroscope{}}) and accelerometer (\textbf{\AccuracyMinAccelerometer{}-\AccuracyMaxAccelerometer{}}) data depending on the history size when conducting four realistic observation sizes. The EER tends to be reduced mainly by a larger observation size and slightly due to an increased history size. This is confirmed by the remaining results: for the magnetometer data we found an EER range of \textbf{\AccuracyMinMagnetometer{}-\AccuracyMaxMagnetometer{}}, for the touchscreen swipes a range of \textbf{\AccuracyMinSwipes{}-\AccuracyMaxSwipes{}}, and for the request logs a range of \textbf{\AccuracyMinBackend{}-\AccuracyMaxBackend{}}. Still, the advantage of larger observations is smaller than it would be expected for two reasons. First, fewer training samples are available as increased observation sizes result in less investigation windows per slice. Second, the same model architecture is used for all sizes, which may be insufficient for larger inputs.
However, an advantage of a bigger observation size is that fewer inferences have to be performed, which are costly due to homomorphic encryption.

These insights answer RQ-1.1: comparing the achieved EERs to related work~\cite{acien2020smartphone}, all considered data sources are suitable for behavior-based authentication, although some approaches accomplish lower values. This is partly due to the more complex scenario and the fact that we tested our models on users that were not part of the training process. Furthermore, recall that the applicable ML models are limited by the homomorphic encryption framework. In previous work, we have shown that lower EER values can be obtained when more sophisticated authentication models are used for unencrypted touchscreen motions~\cite{monschein2021a}. We therefore expect that the results will improve as homomorphic encryption evolves.

Next, we focused on the accuracy of the risk estimation on whole slices, depending on the scenario considered. We vary the setting described in Section~\ref{Sec:Evaluation:Setup:Experiment} as follows: in 50\% of the slices, the behavior of a non-legitimate user was injected (invalid slices). For each invalid slice, the EP was set to either 0\%, 33\%, or 66\% of the slice's size. When calculating the risk level, all combinations of detectors were examined.

To limit the extent of the results, we fixed the size of the history and the observation for each data source. We decided to use values which fall in the middle of the previously considered ranges. Therefore, the history size ($h_{size}$) was set to 30 and the observation size ($o_{size}$) to 7 for all data sources. We measured the EER values depending on the attacker's entry point (EP) and the data sources consulted. In order to keep the table manageable, we have reduced all possible combinations of data sources to the respective number of data sources considered and calculated the average over the values of a group. Table~\ref{Table:Evaluation:Accuracy:Continuous} summarizes the results. Note that the overall EER given includes slices that do not involve an attack scenario, which is why the value does not equal the mean of the three preceding columns.

\input{tables/evaluation/metrics_slices_datasources}

The results can be used to answer RQ-1.2: by combining multiple detectors operating on different data sources, remarkable improvements of the EER values (\textbf{\EERMinOverall{}-\EERMaxOverall{}}) can be realized in our scenario. Consequently, we conclude that the architecture of mPSAuth is suitable for use in continuous authentication environments.

\subsection{Authentication Performance}
We used five platforms to assess performance characteristics (one laptop and four smartphones). The laptop (P1) has 16 GB RAM and a Intel Core i5 CPU with four cores. The smartphones used are common devices: a Galaxy S21 (P2), a Google Pixel 3 (P3), an iPhone 13 (P4) and an iPhone 7 (P5).

We first examine the effort required for homomorphic encryption of the monitoring data. This task must be done mainly by the frontend within the authentication protocol. To estimate the effort required, we measured the actual time needed to encrypt data arising in a single slice (5 minutes) on all platforms. In the experiment, we used the browser-based version of SEAL\footnote{\url{https://github.com/morfix-io/node-seal}} to establish cross-platform comparability. Table~\ref{Table:Evaluation:Performance:Encryption} shows the aggregated results in milliseconds (ms). It lists the first quartile (Q1), the median (Q2), the third quartile (Q3), the mean ($\mu$) and the standard deviation ($\sigma$).
\input{tables/evaluation/performance_encryption}

It can be seen that the results differ from device to device. The iPhone 7 (P5) takes the longest, consuming around \EncryptionMaxTimeMean{} on average to encrypt the data. So in terms of time, encryption takes less than 1\% of the slice in all cases. Furthermore, the encryption can be distributed over time and performed in the background to minimize the impact on the application under observation. Using the results, we can answer RQ-2.1: the time required for encrypting data that accumulates in a 5-minute frame is negligible, even for devices with limited resources.

Next, we quantified the size of the data that the frontend has to send to and receive from the authentication server. We measured both the size of the raw data and the homomorphically encrypted data. Looking at the encrypted data, we found an average data size of \textbf{\EncryptionDataOverheadAvg{}~MB} (minimum: \textit{\EncryptionDataOverheadMin{}~MB}, maximum: \textit{\EncryptionDataOverheadMax{}~MB}), whereas the raw data averaged only \textbf{\RawDataOverheadAvg{}~MB} (minimum: \textit{\RawDataOverheadMin{}~MB}, maximum: \textit{\RawDataOverheadMax{}~MB}). The results correspond to a ratio of \textit{\EncryptionDataOverheadRatio{}x}, by which the encrypted data is larger than the raw data on average. Nevertheless, when looking at the absolute values, the data volumes are reasonable, even for transmission over a mobile network. Note that there is some overhead for verifiable decryption. Since only one additional communication is required and the volume of the exchanged data is limited for proof of correctness~\cite{Fucai2018VerDec}, we neglected the associated overhead. Also, we did not analyze the data amounts sent and received by the backend, as it usually has a powerful network connection. With the collected measurements, we can answer RQ-2.2: The network traffic involved in the authentication protocol is significantly increased (\textit{\EncryptionDataOverheadRatio{}x}) compared to behavior-based authentication atop unencrypted data,  but is nonetheless tolerable.

Finally, we examined how long the required model inferences take per slice for the homomorphically encrypted data. Here, we have only included P1, since the model inferences are performed by the authentication server, which usually has sufficient resources available. Figure~\ref{Fig:Evaluation:Performance:Inference} shows the results as a cumulative distribution function (CDF), which describes the percentage of slices for which a given total execution time for the inferences is not exceeded.

\input{figures/evaluation/performance/inference}

Based on these outcomes, RQ-2.3 can be answered. It can be seen that the inferences on the encrypted data take a considerable amount of time. On average, the processing of one slice requires \textbf{\InferenceTimeEnc{}}, whereas the same inferences on the raw data only need \textbf{\InferenceTimeRaw{}} on average. This corresponds to an \textbf{\InferenceTimeRatio{}-fold} overhead incurred by the adoption of homomorphic encryption. However, these inference times do not hinder the applicability of mPSAuth, as they are remarkably shorter than the duration of a slice (5 minutes). In practice, better performance can be expected for two reasons. First, an authentication server will likely run on much more powerful hardware than P1. Second, recent work in the area of ML in conjunction with homomorphic encryption shows clear potential in improving performance~\cite{FHEFuture1, CheetahSpeedup2021, zhai2021accelerating}.

%% file: figures/evaluation/accuracy/eer_datasources_reduced_split.tex
\usetikzlibrary{
    calc,
    matrix
}
    \newenvironment{customlegend}[1][]{%
        \begingroup
        \csname pgfplots@init@cleared@structures\endcsname
        \pgfplotsset{#1}%
    }{%
        \csname pgfplots@createlegend\endcsname
        \endgroup
    }%
    \def\addlegendimage{\csname pgfplots@addlegendimage\endcsname}
\begin{figure}[!htb]
\centering 
\subfloat[Gyroscope]{
\input{figures/evaluation/accuracy/sub/eer_gyro}
}
\subfloat[Accelerometer]{
\input{figures/evaluation/accuracy/sub/eer_accel}
}

\vspace{0.1cm}

\centering
\begin{tikzpicture}
\begin{customlegend}[legend columns=4,legend style={align=center,column sep=0ex},
        legend entries={$o_{size} = 3$,
                        $o_{size} = 5$,
                        $o_{size} = 7$,
                        $o_{size} = 10$
                        }]
        \addlegendimage{mark=*,dashed,line legend,red}
        \addlegendimage{mark=square*,dashed,line legend,blue}
        \addlegendimage{mark=triangle*,dashed,line legend,orange},
        \addlegendimage{mark=diamond*,dashed,line legend,purple}
        \end{customlegend}
\end{tikzpicture}

\caption{EERs for a single inference depending on the history size ($h_{size}$) and the observation size ($o_{size}$)}
\label{Fig:Evaluation:Accuracy:Single}
\end{figure}
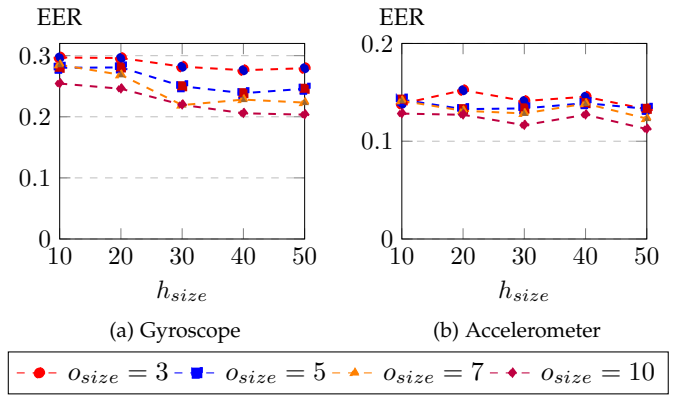

%% file: figures/evaluation/accuracy/sub/eer_gyro.tex
\resizebox{0.48\linewidth}{!}{%
\begin{tikzpicture}
\begin{axis}[
    xlabel={$h_{size}$},
    ylabel={EER},
    xmin=10, xmax=50,
    ymin=0, ymax=0.32,
    xtick={10, 20, 30, 40, 50},
    ytick={0, 0.1, 0.2, 0.3},
    ymajorgrids=true,
    grid style=dashed, width=5cm,
    every axis y label/.style={
        at={(ticklabel* cs:1.05)},
        anchor=south,
    }
]

\addplot+[color=red,mark=*,thick,dashed] coordinates {(10,0.2969900071620941)(20,0.29613474011421204)(30,0.28199300169944763)(40,0.2762092053890228)(50,0.27948116302490234)};
\addplot+[color=blue,mark=square*,thick,dashed] coordinates {(10,0.2801774740219116)(20,0.28073224425315857)(30,0.2499564290046692)(40,0.238600492477417)(50,0.2458711862564087)};
\addplot+[color=orange,mark=triangle*,thick,dashed] coordinates {(10,0.285068154335022)(20,0.26864832639694214)(30,0.21915821731090546)(40,0.2281072437763214)(50,0.2231977880001068)};
\addplot+[color=purple,mark=diamond*,thick,dashed] coordinates {(10,0.2543783187866211)(20,0.245835080742836)(30,0.21955884993076324)(40,0.20573577284812927)(50,0.2034681886434555)};

\end{axis}

\end{tikzpicture}
}

%% file: figures/evaluation/accuracy/sub/eer_accel.tex
\resizebox{0.48\linewidth}{!}{%
\begin{tikzpicture}
\begin{axis}[
    xlabel={$h_{size}$},
    ylabel={EER},
    xmin=10, xmax=50,
    ymin=0, ymax=0.2,
    xtick={10, 20, 30, 40, 50},
    ytick={0, 0.1, 0.2},
    ymajorgrids=true,
    grid style=dashed, width=5cm,
    every axis y label/.style={
        at={(ticklabel* cs:1.05)},
        anchor=south,
    }
]

\addplot+[color=red,mark=*,thick,dashed] coordinates {(10,0.13829530775547028)(20,0.15218578279018402)(30,0.1411350965499878)(40,0.14547917246818542)(50,0.1326238512992859)};
\addplot+[color=blue,mark=square*,thick,dashed] coordinates {(10,0.14255377650260925)(20,0.13279864192008972)(30,0.1335756778717041)(40,0.1389532834291458)(50,0.13313010334968567)};
\addplot+[color=orange,mark=triangle*,thick,dashed] coordinates {(10,0.14135481417179108)(20,0.13114020228385925)(30,0.12841567397117615)(40,0.13848741352558136)(50,0.12315133213996887)};
\addplot+[color=purple,mark=diamond*,thick,dashed] coordinates {(10,0.12825597822666168)(20,0.12703430652618408)(30,0.11642229557037354)(40,0.12708762288093567)(50,0.1124381572008133)};

\end{axis}
\end{tikzpicture}
}

%% file: tables/evaluation/metrics_slices_datasources.tex
\begin{table}[!htb]
    \centering
    \caption{EERs for whole slices depending on the attack scenario and the number of considered data sources}
    
    \begin{tabular}{|c||c|c|c|c|}\hline
        %head
         & \multicolumn{4}{|c|}{\textbf{Attacker Entry Point (EP)}} \\ \hline
        \#Data Sources & 0\% & 33\% & 66\% & (Overall)\\ \hhline{|=|=|=|=|=|}
        % body
        1 & 9.28\% & 10.91\% & 19.22\% & \textbf{14.19\%}\\    \hline
2 & 6.5\% & 7.77\% & 15.94\% & \textbf{11.72\%}\\    \hline
3 & 5.69\% & 6.74\% & 14.88\% & \textbf{10.92\%}\\    \hline
4 & 5.5\% & 6.55\% & 14.81\% & \textbf{10.32\%}\\    \hline
5 & 5.16\% & 6.28\% & 14.8\% & \textbf{9.57\%}\\    \hline
    \end{tabular}
    
    \label{Table:Evaluation:Accuracy:Continuous}
\end{table}

% Min/Max Values as Macros
\newcommand{\EERMaxOverall}{14.19\%}
\newcommand{\EERMinOverall}{9.57\%}

%% file: tables/evaluation/performance_encryption.tex
\begin{table}[!htb]
    \centering
    \caption{Times needed by the different devices to encrypt the behavioral data arising within a 5-minute time frame}
    \begin{tabular}{| c | c | c | c | c | c |}
        \hline
        \textbf{Device} & \textbf{Q1} & \textbf{Q2} & \textbf{Q3} & \textbf{$\boldsymbol{\mu}$} & \textbf{$\boldsymbol{\sigma}$} \\ \hline
        P1 & 42.1ms & 65.5ms & 127.5ms & 121.6ms & 155.6ms \\ \hline
        P2 & 49.4ms & 83.5ms & 152.6ms & 147.3ms & 186.7ms \\ \hline
        P3 & 137.1ms & 244.2ms & 436.1ms & 412.1ms & 515.2ms \\ \hline
        P4 & 61ms & 130ms & 224.5ms & 213.7ms & 267.8ms \\ \hline
        P5 & 139.5ms & 326ms & 555.5ms & 534.2ms & 678.7ms \\ \hline
    \end{tabular}
    \label{Table:Evaluation:Performance:Encryption}
\end{table}

% Max Mean as Macro
\newcommand{\EncryptionMaxTimeMean}{534.2ms}

%% file: figures/evaluation/performance/inference.tex
\begin{figure}[!htb]
\centering
\begin{tikzpicture}
\begin{axis}[
    xlabel={Execution time in seconds (s)},
    ylabel={\% of slices},
    xmin=0, xmax=40,
    ymin=0, ymax=1,
    xtick={0, 10, 20, 30, 40},
    ytick={0.2, 0.4, 0.6, 0.8, 1.0},
    legend columns=2, 
    legend style={
        at={(0.5,-0.3)},
        anchor=north,
        legend columns=2
      },
    ymajorgrids=true,
    grid style=dashed,height=5.1cm
]

\addplot+[
    color=red,
    mark=point,
    thick
    ]
    coordinates {
    % generated via python script
    (0.8,0)(0.81928347123957,0.1028132803286712)(0.9139192769964676,0.1955710375715352)(1.0778385539929352,0.22816621050012442)(1.2417578309894028,0.2500622045284897)(1.4056771079858703,0.26822592684747454)(1.569596384982338,0.28215974122916154)(1.7335156619788057,0.3052998258273203)(1.8974349389752732,0.32296591191838775)(2.0613542159717406,0.34212490669320733)(2.2252734929682085,0.36800199054491173)(2.389192769964676,0.3911420751430705)(2.553112046961144,0.41104752425976615)(2.7170313239576114,0.4374222443393879)(2.880950600954079,0.4630505100771336)(3.0448698779505463,0.4809654142821597)(3.208789154947014,0.5180393132620055)(3.3727084319434817,0.5419258522020403)(3.536627708939949,0.5645683005722816)(3.700546985936417,0.5872107489425229)(3.8644662629328845,0.608360288629012)(4.028385539929352,0.6267728290619555)(4.1923048169258195,0.6484200049763621)(4.356224093922288,0.6678278178651404)(4.520143370918754,0.6810151779049514)(4.684062647915223,0.695197810400597)(4.84798192491169,0.7076387160985318)(5.011901201908158,0.7185867131127144)(5.175820478904625,0.7330181637223188)(5.339739755901093,0.7397362527992035)(5.503659032897561,0.7464543418760883)(5.667578309894028,0.7506842498133861)(5.831497586890496,0.7561582483204774)(5.995416863886963,0.7616322468275687)(6.159336140883431,0.7673550634486187)(6.323255417879898,0.7695944264742469)(6.487174694876366,0.7723314257277926)(6.651093971872834,0.7765613336650904)(6.815013248869302,0.7788006966907186)(6.978932525865769,0.7822841502861404)(7.1428518028622365,0.7845235133117686)(7.306771079858704,0.7867628763373968)(7.4706903568551715,0.7892510574769838)(7.63460963385164,0.7934809654142816)(7.798528910848107,0.7957203284399098)(7.962448187844575,0.7987061458074142)(8.126367464841042,0.8019407812888772)(8.290286741837509,0.8056730529982576)(8.454206018833977,0.809405324707638)(8.618125295830446,0.8133864145309772)(8.782044572826912,0.816869868126399)(8.94596384982338,0.8210997760636968)(9.109883126819847,0.8245832296591186)(9.273802403816315,0.8275690470266229)(9.437721680812784,0.8303060462801686)(9.60164095780925,0.834038317989549)(9.765560234805719,0.8382682259268468)(9.929479511802185,0.8437422244339381)(10.093398788798654,0.8477233142572772)(10.257318065795122,0.8529484946504098)(10.421237342791589,0.857676038815625)(10.585156619788057,0.8638964916645925)(10.749075896784523,0.8696193082856425)(10.912995173780992,0.8743468524508577)(11.076914450777458,0.8818113958696187)(11.240833727773927,0.8885294849465034)(11.404753004770395,0.8942523015675534)(11.568672281766862,0.8979845732769338)(11.73259155876333,0.9014680268723556)(11.896510835759797,0.9034585717840252)(12.060430112756265,0.9059467529236122)(12.224349389752732,0.9106742970888274)(12.3882686667492,0.9129136601144556)(12.552187943745668,0.9163971137098774)(12.716107220742135,0.9191341129634231)(12.880026497738603,0.9206270216471752)(13.04394577473507,0.9238616571286383)(13.207865051731538,0.9273451107240601)(13.371784328728006,0.9325702911171927)(13.535703605724473,0.9370490171684492)(13.699622882720941,0.9387907439661601)(13.863542159717408,0.9417765613336645)(14.027461436713876,0.9435182881313754)(14.191380713710343,0.9452600149290863)(14.355299990706811,0.9467529236128385)(14.51921926770328,0.9472505598407559)(14.683138544699746,0.9479970141826319)(14.847057821696215,0.9484946504105494)(15.010977098692681,0.9492411047524254)(15.17489637568915,0.9497387409803428)(15.338815652685616,0.9502363772082603)(15.502734929682084,0.9517292858920124)(15.666654206678553,0.9517292858920124)(15.83057348367502,0.9519781040059712)(15.994492760671488,0.9527245583478472)(16.158412037667954,0.9534710126897232)(16.322331314664424,0.953719830803682)(16.48625059166089,0.954466285145558)(16.650169868657358,0.955212739487434)(16.814089145653824,0.9562080119432688)(16.978008422650294,0.9569544662851448)(17.14192769964676,0.9574521025130622)(17.305846976643227,0.9579497387409797)(17.469766253639694,0.9581985568549384)(17.633685530636164,0.9581985568549384)(17.79760480763263,0.9589450111968144)(17.961524084629097,0.9589450111968144)(18.125443361625567,0.9589450111968144)(18.289362638622034,0.9591938293107731)(18.4532819156185,0.9591938293107731)(18.61720119261497,0.9591938293107731)(18.781120469611437,0.9604379198805666)(18.945039746607904,0.9606867379945253)(19.10895902360437,0.9611843742224427)(19.27287830060084,0.9614331923364015)(19.436797577597307,0.9616820104503602)(19.600716854593774,0.9626772829061949)(19.764636131590244,0.9629261010201536)(19.92855540858671,0.9631749191341124)(20.092474685583177,0.9636725553620298)(20.256393962579644,0.9639213734759885)(20.420313239576114,0.9644190097039059)(20.58423251657258,0.9649166459318234)(20.748151793569047,0.9651654640457821)(20.912071070565517,0.9654142821597408)(21.075990347561984,0.9654142821597408)(21.23990962455845,0.9656631002736995)(21.403828901554917,0.9656631002736995)(21.567748178551387,0.9656631002736995)(21.731667455547854,0.9656631002736995)(21.89558673254432,0.9656631002736995)(22.05950600954079,0.9656631002736995)(22.223425286537257,0.9659119183876582)(22.387344563533723,0.9661607365016169)(22.55126384053019,0.9664095546155756)(22.71518311752666,0.9664095546155756)(22.879102394523127,0.9669071908434931)(23.043021671519593,0.9669071908434931)(23.206940948516063,0.9669071908434931)(23.37086022551253,0.9671560089574518)(23.534779502508997,0.9674048270714105)(23.698698779505463,0.9676536451853692)(23.862618056501933,0.9679024632993279)(24.0265373334984,0.9684000995272454)(24.190456610494866,0.9688977357551628)(24.354375887491337,0.9691465538691215)(24.518295164487803,0.9691465538691215)(24.68221444148427,0.9691465538691215)(24.846133718480736,0.9693953719830802)(25.010052995477206,0.969644190097039)(25.173972272473673,0.969644190097039)(25.33789154947014,0.9698930082109977)(25.50181082646661,0.9701418263249564)(25.665730103463076,0.9701418263249564)(25.829649380459543,0.9701418263249564)(25.993568657456013,0.9706394625528738)(26.15748793445248,0.9708882806668325)(26.321407211448946,0.9708882806668325)(26.485326488445413,0.9711370987807912)(26.649245765441883,0.9711370987807912)(26.81316504243835,0.97138591689475)(26.977084319434816,0.9716347350087087)(27.141003596431286,0.9721323712366261)(27.304922873427753,0.9726300074645435)(27.46884215042422,0.9726300074645435)(27.632761427420686,0.9736252799203783)(27.796680704417156,0.9736252799203783)(27.960599981413623,0.9736252799203783)(28.12451925841009,0.973874098034337)(28.28843853540656,0.973874098034337)(28.452357812403026,0.973874098034337)(28.616277089399492,0.9741229161482957)(28.78019636639596,0.9741229161482957)(28.94411564339243,0.9743717342622544)(29.108034920388896,0.9746205523762131)(29.271954197385362,0.9748693704901719)(29.435873474381832,0.9748693704901719)(29.5997927513783,0.9748693704901719)(29.763712028374766,0.9751181886041306)(29.927631305371232,0.9753670067180893)(30.091550582367702,0.9753670067180893)(30.25546985936417,0.9753670067180893)(30.419389136360635,0.9753670067180893)(30.583308413357106,0.9753670067180893)(30.747227690353572,0.9753670067180893)(30.91114696735004,0.975615824832048)(31.075066244346505,0.975615824832048)(31.238985521342975,0.975615824832048)(31.402904798339442,0.9758646429460067)(31.56682407533591,0.9763622791739242)(31.73074335233238,0.9768599154018416)(31.894662629328845,0.977357551629759)(32.058581906325315,0.978104005971635)(32.22250118332178,0.9783528240855938)(32.38642046031825,0.9783528240855938)(32.550339737314715,0.9795969146553872)(32.71425901431118,0.979845732769346)(32.87817829130765,0.979845732769346)(33.042097568304115,0.9803433689972634)(33.20601684530059,0.9805921871112221)(33.369936122297055,0.9805921871112221)(33.53385539929352,0.9805921871112221)(33.69777467628999,0.9808410052251808)(33.861693953286455,0.9808410052251808)(34.02561323028292,0.9810898233391395)(34.18953250727939,0.9810898233391395)(34.35345178427586,0.981587459567057)(34.51737106127233,0.9818362776810157)(34.681290338268795,0.9820850957949744)(34.84520961526526,0.9825827320228918)(35.00912889226173,0.9825827320228918)(35.173048169258195,0.9833291863647678)(35.33696744625466,0.9838268225926853)(35.500886723251135,0.984075640706644)(35.6648060002476,0.98482209504852)(35.82872527724407,0.9850709131624787)(35.992644554240535,0.9853197312764375)(36.156563831237,0.9858173675043549)(36.32048310823347,0.9860661856183136)(36.48440238522994,0.9863150037322723)(36.64832166222641,0.9863150037322723)(36.812240939222875,0.9868126399601898)(36.97616021621934,0.9873102761881072)(37.14007949321581,0.9878079124160246)(37.303998770212274,0.988305548643942)(37.46791804720874,0.9890520029858181)(37.631837324205215,0.9893008210997768)(37.79575660120168,0.9895496392137355)(37.95967587819815,0.9897984573276942)(38.123595155194614,0.9900472754416529)(38.28751443219108,0.9910425478974877)(38.45143370918755,0.9910425478974877)(38.615352986184014,0.9915401841254051)(38.77927226318049,0.9917890022393638)(38.943191540176954,0.9920378203533226)(39.10711081717342,0.9922866384672813)(39.27103009416989,0.9927842746951987)(39.434949371166354,0.9932819109231161)(39.59886864816282,0.9935307290370748)(39.76278792515929,0.9937795471510336)(39.92670720215576,0.9937795471510336)(40.09062647915223,0.9940283652649923)(40.254545756148694,0.994277183378951)(40.41846503314516,0.9947748196068684)(40.58238431014163,0.9950236377208271)(40.746303587138094,0.9952724558347859)(40.91022286413456,0.9957700920627033)(41.074142141131034,0.996018910176662)(41.2380614181275,0.9965165464045794)(41.40198069512397,0.9967653645185381)(41.565899972120434,0.9972630007464556)(41.7298192491169,0.997760636974373)(41.89373852611337,0.998507091316249)(42.057657803109834,0.9990047275441665)(42.22157708010631,0.9990047275441665)(42.385496357102774,0.9990047275441665)(42.54941563409924,0.9992535456581252)(42.71333491109571,0.9992535456581252)(42.877254188092174,0.9995023637720839)(43.04117346508864,0.9995023637720839)(43.20509274208511,0.9995023637720839)(43.36901201908158,0.9995023637720839)(43.53293129607805,0.9997511818860426)(43.696850573074514,0.9997511818860426)(43.86076985007098,0.9997511818860426)(44.02468912706745,0.9997511818860426)(44.18860840406391,0.9997511818860426)(44.35252768106038,0.9997511818860426)(44.516446958056854,0.9997511818860426)(44.68036623505332,0.9997511818860426)(44.84428551204979,0.9997511818860426)(45.00820478904625,0.9997511818860426)(45.17212406604272,0.9997511818860426)(45.33604334303919,0.9997511818860426)(45.49996262003565,0.9997511818860426)(45.66388189703213,0.9997511818860426)(45.82780117402859,0.9997511818860426)(45.99172045102506,0.9997511818860426)(46.15563972802153,0.9997511818860426)(46.31955900501799,0.9997511818860426)(46.48347828201446,0.9997511818860426)(46.647397559010926,0.9997511818860426)(46.8113168360074,0.9997511818860426)(46.97523611300387,0.9997511818860426)(47.13915539000033,0.9997511818860426)(47.3030746669968,0.9997511818860426)(47.466993943993266,0.9997511818860426)(47.63091322098973,0.9997511818860426)(47.7948324979862,0.9997511818860426)(47.95875177498267,0.9997511818860426)(48.12267105197914,0.9997511818860426)(48.286590328975606,0.9997511818860426)(48.45050960597207,0.9997511818860426)(48.61442888296854,0.9997511818860426)(48.778348159965006,0.9997511818860426)(48.94226743696147,0.9997511818860426)(49.106186713957946,0.9997511818860426)(49.27010599095441,0.9997511818860426)(49.43402526795088,0.9997511818860426)(49.597944544947346,0.9997511818860426)(49.76186382194381,0.9997511818860426)(49.92578309894028,1.0)
    };

\end{axis}
\end{tikzpicture}
\caption{Execution times for inferences on homomorphically encrypted data within one slice}
\label{Fig:Evaluation:Performance:Inference}
\end{figure}
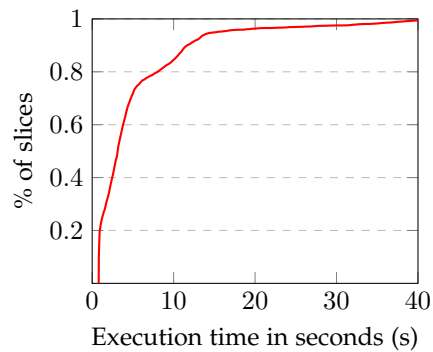

%% file: sections/relatedwork.tex
\section{Related Work}
\label{Sec:RelatedWork}
Related work in the area of behavior-based authentication can be divided into approaches that work with data that originates in the frontend or the backend of the application.

% Block Frontend - No Privacy
The group of approaches that focus on the frontend can be subdivided depending on whether or not they incorporate privacy protection mechanisms. As examples for approaches without privacy protection, Karanikiotis et al.~\cite{Karanikiotis2020TouchTraces} analyzed touch traces and trained user-specific models based on Support Vector Machines (SVM) to authenticate users continuously. FAR and FRR of less than 5\% were measured for realistic scenarios. MultiLock~\cite{acien2019multilock} is an approach that leverages data sources for authentication on mobile devices, including sensors, touchscreen interactions, location, and network information. Depending on the scenario, EER values of well below 5\% were determined. In addition, many other efforts deal with behavior-based authentication based on various data sources, such as location data~\cite{thao2020gps, PengGlasses2017}, sensor data~\cite{SCANetSensorAuth2020, LiSensorAuth2018}, and network connection information~\cite{LiNetworkAuth2018}. Besides, Acien et al.~\cite{acien2020smartphone} summarized further work in this area in a structured way. However, even though many of these works achieve single-digit EERs, compared to mPSAuth, they are mainly intended for authentication directly on the mobile device and therefore not suitable for use in mobile web applications. For this reason, they lack a profound protection of privacy. Moreover, \cite{thao2020gps, acien2019multilock, Karanikiotis2020TouchTraces} train user-specific models, which limits their scalability.

% Block Frontend + Privacy
Govindarajan et al.~\cite{TouchTracesPrivacy2013} developed an authentication protocol that uses touchscreen swipes as data source while considering privacy protection based on homomorphic encryption. The EER ranges from 22\% to 38\% and is comparable to the values we achieved for the sole analysis of swipes. In contrast to mPSAuth, the work adopts an honest-but-curious attacker model~\cite{paverd2014modelling}, where the attacker is assumed to comply with the protocol. For mobile web applications, this is not sufficient. Research by Vassallo et al.~\cite{VassalloPrivacyAuth2017} and Sun et al.~\cite{SunTouchscreenAuthRemovePrivacy2015} also uses touchscreen interactions, despite taking a different approach to privacy protection. They transform the data so that sensitive information is removed or replaced. These methods cannot provide security guarantees, which is possible with homomorphic encryption. Furthermore, the methodology of Vassallo et al.~\cite{VassalloPrivacyAuth2017} indicates that the accuracy of the authentication may be significantly reduced by the application of privacy-protecting measures. There are procedures based on cancelable biometrics~\cite{PatelCancelable2015} that transform the behavioral data for privacy protection while it remains comparable with baseline data. Hatin et al.~\cite{Hatin2017PrivacyPT} use the \enquote{BioHashing} technique, a subtype of cancelable biometrics, to authenticate users via their communication behavior (phone calls and messages). Attacks on BioHashing~\cite{Topcu2016, LacharmeBioHashingAttack2013} reveal weaker security than with homomorphic encryption. Other privacy-preserving approaches to user authentication are summarized in~\cite{PrivacySummary2021}.

% Block Backend
The last group of related work aims to authenticate users based on information available on the backend. Freeman, Jain et al.~\cite{Freeman2016WhoAY} presented a strategy to detect illegitimate users based on request parameters such as IP address and browser configuration. In an extensive evaluation using real-world data from a social network, they reported an FPR of 10\%. Ongun et al.~\cite{Ongun2019AuthIoT} describe a similar approach operating on network traffic (e.g., timestamps and packet lengths) to authenticate users of home-based internet of things devices (accuracy of 86\%-97\%). Other works that address user authentication based on backend information are summarized in~\cite{AlacaDeviceFingerprint2016}. Due to the focus on the backend, many types of attacks cannot be detected (e.g. device theft). Moreover, we are not aware of any approach that involves privacy protection, as the authentication server and the backend often coincide in practice. In contrast, with privacy protection in mPSAuth it is possible to support scenarios in which the authentication is provided as an external service.

%% file: sections/conclusion.tex
\section{Conclusion and Future Work}
\label{Sec:Conclusion}
In this paper, we proposed mPSAuth, an approach to continuously authenticate users of mobile web applications based on behavioral data collected during usage. mPSAuth relies on an authentication protocol that employs homomorphic encryption to ensure that the user's sensitive behavioral data is not disclosed in the authentication process. The behavioral data is gathered from both the frontend (e.g., touchscreen interactions and sensor data) and the backend (e.g., request logs) of the application under consideration. Finally, the homomorphically encrypted data is analyzed using machine learning techniques to determine a risk level that reflects the likelihood that a present user is an illegitimate person. mPSAuth is designed for scalability, as only a fixed number of machine learning models need to be trained and then applied to all users. The contributions of mPSAuth address challenges that have been inadequately addressed by existing approaches. In particular, this concerns the protection of privacy and the scalable analysis of user behavior based on machine learning techniques.

We evaluated mPSAuth in an extensive experiment designed around a publicly available dataset collected in a mobile gaming application. We analyzed the accuracy of authentication using different data sources and determined EER values in the range of \AccuracyMinAccelerometer{}-\AccuracyMaxMagnetometer{}. Subsequently, we showed that combining data sources can reduce the EER rates down to \EERMinOverall{}. These values confirmed the suitability of mPSAuth for behavior-based authentication, even though they are slightly higher compared to some related work. The main reason for this is the complex scenario considered and the limitations for building the machine learning models inherent to the available homomorphic encryption frameworks. Furthermore, the computational overhead for encrypting the monitoring data and the inference on encrypted data was measured, as well as the network traffic generated by the authentication protocol. We found a moderate increase in traffic as well as a significant increase in inference times. Nevertheless, the absolute values indicated that the approach can still be applied in practice.

There is room for improvements and extensions in future work. In particular, this concerns the incorporation of feedback from previous authentication processes as outlined in previous work~\cite{monschein2021a}. In this way, the authentication system could adapt to users in the long run and achieve higher accuracy. Moreover, we intend to extend mPSAuth so that incidences where an attacker has knowledge about the behavior of other users (e.g., replay attacks) can be mitigated efficiently. Besides, there is potential in terms of the performance of mPSAuth. As performance mainly depends on the underlying homomorphic encryption scheme and a lot of research is conducted in this area~\cite{FHEFuture1, CheetahSpeedup2021, zhai2021accelerating}, there is great potential for our approach to become more resource-efficient. In addition, more sophisticated models could be used, which could improve the accuracy. Finally, we plan a end-to-end evaluation within a real-world environment to validate the general applicability of our approach.